\documentclass{article} 
\usepackage{amssymb,amsthm,amsmath,amsfonts} 
\usepackage{epsfig}

\renewcommand{\u}{y}
\renewcommand{\v}{z}
\newcommand{\w}{w}

\begin{document}

\title{An invariant in shock clustering and Burgers turbulence} 
\author{Ravi Srinivasan\textsuperscript{1}}

\date{\today}

\maketitle

\begin{abstract}
1-D scalar conservation laws with convex flux and Markov initial data are now known to yield a completely integrable Hamiltonian system. In this article, we rederive the analogue of Loitsiansky's invariant in hydrodynamic turbulence from the perspective of integrable systems. Other relevant physical notions such as energy dissipation and spectrum are also discussed. 
\end{abstract}



\medskip 
\noindent
\footnotetext[1] {Department of Mathematics, The University of Texas at Austin,
Austin, TX 78712.
Email: rav@math.utexas.edu}

\section{Introduction}

The inviscid Burgers equation
\begin{equation}
	\partial_{t}u+
	\partial_{x}\left(\frac{u^{2}}{2}\right)=0,\qquad x\in\mathbb{R},t>0,\qquad u(0,x)=u_{0}(x)\label{eq:Burgers}
\end{equation}
with random initial data $u_{0}$ has been extensively studied over the course of several decades. As a caricature of hydrodynamic turbulence, the earliest fundamental results are due to Burgers himself, who studied (\ref{eq:Burgers}) for $u_{0}$ a white noise process in space \cite{Burgers}. When considered with an added random forcing term, the equation becomes a model of random growth with deposition. Taken together, these form the basis of Burgers-KPZ turbulence. A broad overview of these areas can be found in \cite{Woc} and the review article \cite{Bec}.

Many essential features of Burgers turbulence are preserved in the more general context of 1-D scalar conservation laws with strictly convex flux and random initial data. In recent work \cite{MS1}, we have shown the class of Markov processes in space with only downward jumps is preserved by the deterministic dynamics. Although it is one of the simplest examples of a nontrivial flow on the space of probability measures, this model has surprisingly rich behavior. The evolution is given in terms of a Lax equation, and it has deep connections to kinetic theory, random matrices, and statistics. Moreover, it has been shown to be a completely integrable Hamiltonian system \cite{M1}. As an integrable system, its evolution is constrained by conserved quantities. In this article we examine how one particular conserved quantity of physical interest, well-known in the turbulence literature, arises in our setting.

\subsection{Entropy solutions to scalar conservation laws}

To begin, let us review some basic facts about the 1-D scalar conservation law
\begin{equation}
	\partial_{t}u+
	\partial_{x}f(u)=0,\qquad x\in\mathbb{R},t>0\label{eq:ConservationLaw}
\end{equation}
with strictly convex flux $f\in C^{1}$ and initial data $u(x,0)=u_{0}(x)$. Classical solutions found using the method of characteristics typically cease to exist after a finite time due to the formation of shocks. Instead, a suitable global-in-time solution to (\ref{eq:ConservationLaw}) is the \emph{entropy solution}, derived via a variational principle for the action functional 
\begin{equation}
	I(s;x,t)=U_{0}(s)+tf^{*}\left(\frac{x-s}{t}\right).
\end{equation}
Here, $U_{0}(s)=\int_{0}^{s}u_{0}(r)dr$ is the initial potential and $f^{*}(s)=\sup_{u\in\mathbb{R}}\{us-f(u)\}$ the Legendre transform of the flux. We assume that $U_{0}$ has no upward jumps and that $\lim_{|x|\to\infty}I(s;x,t)=+\infty$ always holds, ensuring that the infimum of $I$ is achieved. The \emph{inverse Lagrangian function} $a(x,t)$ is then defined by the Hopf-Lax formula:
\begin{equation}
	a(x,t)=\arg^{+}\min_{s\in\mathbb{R}}I(s;x,t).
\end{equation}
The $^{+}$ denotes that we choose $a(x,t)$ to be the largest value where $I$ achieves its minimum. In terms of infinitesimal particles, $a$ gives the (rightmost) initial location of the particle at location $x$ at time $t$. For fixed $t>0$, $a(x,t)$ is increasing in $x$. At points of continuity of $a(\cdot,t)$ the entropy solution is 
\begin{equation}
	u(x,t)=(f')^{-1}\left(\frac{x-a(x,t)}{t}\right).
\end{equation}
Downward jumps in $u$ correspond to \emph{shocks,} and the velocity of a shock at $x$ with left- and right-limits $u_{\pm}=u(x_{\pm},t)$ satisfies the \emph{Rankine-Hugoniot condition}
\begin{equation}
	u(x,t)=\frac{f(u_{-})-f(u_{+})}{u_{-}-u_{+}}=:[f]_{u_{-},u_{+}}.
\end{equation}

\subsection{Loitsiansky's invariant}

Now suppose that $u_{0}(x)$ is a mean-zero velocity field which is statistically stationary and ergodic in $x$. Then for each $t>0$, the entropy solution $u(x,t)$ remains stationary ergodic and ensemble averages are equivalent to spatial averages (which we denote by brackets). It has long been known that (\ref{eq:Burgers}) admits a conserved quantity analogous to \emph{Loitsiansky's invariant} in hydrodynamic turbulence \cite{L1}:
\begin{equation}
	J_{0}(t)=\int_{0}^{\infty}\langle u(x,t)u(x+h,t)\rangle dh.\label{eq:J0}
\end{equation}
$J_{0}$ is simply the integral of the two-point correlation function, and is essentially the value of the energy spectrum in the limit of low wavenumbers. In a series of articles \cite{B_corra,B_corrb,B_corrc,B_corrd}, Burgers provided three distinct derivations of the invariance of $J_{0}$. Upon doing so he investigated its implications on the evolution of a variety of shock statistics. Intriguingly, many of Burgers' calculations involve (unclosed) kinetic equations describing shock coalescence as in \cite{EvE}, and are quite prescient in that respect.

It is easily seen that $J_{0}$ is preserved for any 1-D scalar conservation law. A formal derivation is as follows. Denoting $u(x)=u(x,t)$, one has that for any $h>0$,
\begin{align*}
	\partial_{t}\langle u(x)u(x+h)\rangle & =\left\langle u(x)
	\partial_{t}u(x+h)+
	\partial_{t}u(x)u(x+h)\right\rangle \\
	& =-\left\langle u(x)
	\partial_{x}f(u(x+h))+
	\partial_{x}f(u(x))u(x+h)\right\rangle .
\end{align*}
Therefore, the two-point correlation function $\theta(h)=\langle u(x)u(x+h)\rangle$ itself satisfies a conservation law with flux $\tau(h)=\left\langle u(x)\left(f(u(x+h))-f(u(x-h))\right)\right\rangle $:
\begin{equation}
	\partial_{t}\theta(h)+
	\partial_{h}\tau(h)=0.\label{eq:CorrelationConservationLaw}
\end{equation}
For Burgers' flux $f(u)=u^{2}/2$, this is the analogue of the K\'arm\'an-Howarth equation for three-dimensional isotropic turbulence in the vanishing viscosity limit. Assuming that the correlation length of the initial velocity field decays sufficiently fast, one has that $\tau(h)\to0$ as $h\to\infty$ and
\[ 
\partial_{t}J_{0}(t)=
\partial_{t}\int_{0}^{\infty}\theta(h)dh=0.\]
If, in addition, $u(x)$ is statistically invariant under the transformation $x\mapsto-x$, $u\mapsto-u$, the correlation flux takes the form
\begin{equation}
	\tau(h)=-2\left\langle f(u(x))u(x+h)\right\rangle .
\end{equation}
When $f(u)=u^{2}/2$, this implies that $\tau=-\frac{1}{6}S_{3}$ with third-order structure function $S_{3}(h)=\left\langle (u(x+h)-u(x))^{3}\right\rangle $. This is particular to Burgers equation since $\tau$ typically cannot be expressed in a simple manner using structure functions.

Finally, we briefly remark on the physical relevance of the conserved quantity (\ref{eq:J0}) for isotropic Navier-Stokes turbulence in the infinite domain. In this context, it is widely believed that certain universal scaling laws, such as a law of energy decay, should hold for the statistics of the velocity field. Dimensional analysis dictates that the mean energy dissipation should be proportional to $U^{3}/L$, where $U(t)$ is the root-mean square velocity and $L(t)$ is the integral length scale. A self-similarity ansatz for the velocity correlation function \cite{vKL} relates the scaling of $U(t)$ and $L(t)$---however, an additional relation is needed to obtain the scaling exponents themselves. The invariance of $J_{0}$, commonly referred to as the {}``permanence of large eddies,'' was utilized by Kolmogorov \cite{Kolm2} to obtain a power law for the energy decay. While this was later shown to hold only under particular assumptions on the initial spectrum, it remains a fundamental result in the theory. In the simpler setting of Burgers turbulence, many of the same issues have been thoroughly investigated and settled. We direct the interested reader to \cite{Bec,Gurbatov,Kida} for a comprehensive discussion.

\subsection{Lax equation and integrability of shock clustering}

To start, we take $u_{0}(x)$ to be a stationary Markov process with only downward jumps. In \cite{MS1}, it was proven that the entropy solution to (\ref{eq:ConservationLaw}) then satisfies a closure property: for each fixed $t>0$, $u(x,t)$ is a stationary Markov process in $x$ with only downward jumps. This was also shown to hold if the initial data is the derivative of a stationary L\'evy process in $x$ with only downward jumps (e.g., white noise). A formal derivation of this closure for Burgers equation was obtained in \cite{CD}.

Denote by $p(d\u,t)$ and $\left\{ q_{h}(\u,d\v,t)\right\} _{h>0}$ the stationary and transition measures in $x$ of the process $u(x,t)$. The evolution of the statistics can be stated in terms of the generator of the solution process. First, define the 1- and 2-point operators through their action on appropriate test functions $\varphi$:
\begin{equation}
	\mathcal{P}(t)\varphi(\u)=\int_{\mathbb{R}}p(d\u,t)\varphi(\u),\qquad\mathcal{Q}_{h}(t)\varphi(\u)=\int_{\mathbb{R}}q_{h}(\u,d\v,t)\varphi(\v).\label{eq:1-2PointOps}
\end{equation}
Consider the generator $\mathcal{A}(t)\varphi=\lim_{h\downarrow0}\frac{1}{h}\left(\mathcal{Q}_{h}(t)\varphi-\varphi\right)$ of the Markov semigroup in $x$ (assuming the Feller property holds). For each fixed $t>0$, it has the form
\begin{equation}
	\mathcal{A}(t)\varphi(\u)=b(\u,t)\varphi'(\u)+\int_{\mathbb{R}}n(\u,d\v,t)(\varphi(\v)-\varphi(\u)),\label{eq:GenA}
\end{equation}
where $b(\u,t)$ is the drift coefficient of the process and $n(\u,d\v,t)$ is the jump measure. Now we define an operator $\mathcal{B}(t)$ which, for each fixed $x$, serves as a `generator' of the solution in $t$: 
\begin{equation}
	\mathcal{B}(t)\varphi(\u)=-b(\u,t)f'(\u)\varphi'(\u)-\int_{\mathbb{R}}n(\u,d\v,t)[f]_{\u,\v}(\varphi(\v)-\varphi(u)).\label{eq:GenB}
\end{equation}
$\mathcal{B}(t)$ truly is the generator of a Markov semigroup if $f$ is decreasing. As shown in \cite{MS1}, the evolution of the 1- and 2-point operators then takes the form
\begin{equation}
	\partial_{t}\mathcal{P}=\mathcal{P}\mathcal{B},\qquad
	\partial_{t}\mathcal{Q}_{h}=[\mathcal{Q}_{h},\mathcal{B}]\label{eq:1-2PointEvolution}
\end{equation}
where square brackets denote the commutator. In terms of generators, this is the Lax equation
\begin{equation}
	\partial_{t}\mathcal{A}=[\mathcal{A},\mathcal{B}].\label{eq:LaxEqn}
\end{equation}
We have shown that (\ref{eq:LaxEqn}) is equivalent to a kinetic equation for the shock statistics. These equations (remarkably!) admit explicit solutions when $u_{0}$ is a Brownian motion \cite{B_burgers,Duchon0,Duchon1,Sinai} or a white noise \cite{Burgers,Frachebourg,Groeneboom}. In addition, there are many intriguing links to random matrices and integrable systems. This is discussed at length in \cite{MS1}.

\section{The Manakov relation and an invariant of the Lax equation\label{sub:DerivationMarkovSemigroup}}

While the Lax equation yields some conserved quantities, these are not sufficient to fully describe the evolution of the system. One can, in addition, verify that the operators $\mathcal{A}$ and $\mathcal{B}$ satisfy the Manakov relation
\begin{equation}
	[\mathcal{A},\mathcal{N}]-[\mathcal{M},\mathcal{B}]=0.\label{eq:Manakov}
\end{equation}
with multiplication operators
\begin{equation}
	\mathcal{M}\varphi(\u)=\u\varphi(\u),\qquad\mathcal{N}\varphi(\u)=f(\u)\varphi(\u).\label{eq:MultiplicationOps}
\end{equation}
Equation (\ref{eq:Manakov}) is rigid in that it only holds with multiplication operators $\mathcal{M}$ and $\mathcal{N}$: if we let $\mathcal{M}_{\psi}\varphi(\u)=\psi(\u)\varphi(\u)$ for any $\psi$, then $[\mathcal{A},\mathcal{M}_{\psi_{A}}]-[\mathcal{M}_{\psi_{B}},\mathcal{B}]=0$ if and only if $\psi_{A}(\u)=f(\u)$ and $\psi_{B}(\u)=\u$. 

The Manakov relation yields the necessary additional integrals by allowing a spectral parameter to be introduced as in \cite{Manakov}. That is, by (\ref{eq:LaxEqn}) and (\ref{eq:Manakov}),
\begin{equation}
	\partial_{t}(\mathcal{A}-\mu\mathcal{M})=[\mathcal{A}-\mu\mathcal{M},\mathcal{B}+\mu\mathcal{N}],\qquad\mu\in\mathbb{C}.
\end{equation}
If one considers discretizations of $\mathcal{A}$ and $\mathcal{B}$ by $N\times N$ matrices, this implies that the spectral curve
\begin{equation}
	\Gamma=\left\{ (\lambda,\mu)\in\mathbb{C}^{2}|\det(\mathcal{A}-\lambda\text{Id}-\mu\mathcal{M})=0\right\} .
\end{equation}
remains unchanged in time. The conserved quantities are then given by the coefficients of the characteristic polynomial. It has recently been shown by Menon \cite{M1} that this discretized system, the Markov $N$-wave model, is a completely integrable Hamiltonian system. While the argument above shows that the spectrum is invariant in the discrete setting, this does not remain true as $N\to\infty$ since the continuous spectrum may evolve. Therefore, it is an interesting and open problem to find conserved quantities that survive in the continuum limit, and to determine which of these are finite. Since the discrete model does not allow for nontrivial statistically stationary solutions unless one takes $N\to\infty$, we note that the invariant $J_{0}$ can only appear in this limit.

Let us now state our main result. In what follows we will assume that $p(d\u,t)$ and $q_{h}(\u,d\v,t)$ decay fast enough for the appropriate integrals to converge. For stationary Markov solutions to (\ref{eq:ConservationLaw}), we find that the invariance of
\begin{equation}
	J_{0}=\int_{0}^{\infty}\mathbb{E}[u(x,t)u(x+h,t)]dh=\int_{0}^{\infty}\left(\int_{\mathbb{R}}p(d\u,t)\u\int_{\mathbb{R}}q_{h}(\u,d\v,t)\v\right)dh\label{eq:J0Markov}
\end{equation}
is equivalent to the identity
\begin{equation}
	\mathcal{P}\mathcal{A}\left(f(\u)\mathcal{Q}_{h}\u\right)=0.\label{eq:CorrelationInvariant}
\end{equation}
Since $u$ is stationary in $x$, this is the weak form of the forward equation $0=\mathcal{A}^{\dagger}p$ with test function $f(\u)\mathcal{Q}_{h}\u$. More generally, fix $n\in\mathbb{N}$, denote $h_{i}=x_{i}-x_{i-1}$ and $\mathcal{Q}_{i}=\mathcal{Q}_{h_{i}}$ for $i=1,\dots,n-1$, and let $\mathbf{h}=(h_{1},\dots,h_{n-1})$. We demonstrate that the Manakov relation (\ref{eq:Manakov}) implies the $n$-point function $\theta(\mathbf{h},t)=\mathcal{P}\u\mathcal{Q}_{1}\u\cdots\mathcal{Q}_{n-1}\u$ satisfies the conservation law
\begin{equation}
	\partial_{t}\theta(\mathbf{h})+\nabla\cdot\mathbf{T}(\mathbf{h})=0\label{eq:MarkovCorrelationConservationLaw}
\end{equation}
with flux $\mathbf{T}_{i}(\mathbf{h},t)=\mathcal{P}\u\mathcal{Q}_{1}\u\cdots\mathcal{Q}_{i-1}(f(\u)\mathcal{Q}_{i}\u-\u\mathcal{Q}_{i}f(\u))\mathcal{Q}_{i+1}\u\cdots\mathcal{Q}_{n-1}\u$. Integration with respect to $\mathbf{h}$ shows that 
\begin{equation}
	J_{0}^{(n)}=\int_{x_{0}<x_{1}<\cdots<x_{n-1}}\mathbb{E}[u(x_{0},t)u(x_{1},t)\cdots u(x_{n-1},t)]dx_{1}\cdots dx_{n-1}\label{eq:J0(n)Markov}
\end{equation}
is conserved by the evolution.

We note that these invariants can also be arrived at by the formal argument in the introduction, as the essential properties are the stationarity of the velocity field and sufficient decay in the correlation length scale. The purpose of this article, however, is not to prove a new result but to provide a novel perspective on the matter from the viewpoint of integrable systems.

\subsection{Derivation of main result}

For simplicity we will suppress $t$ in the notation whenever possible. We begin by showing the equivalence of (\ref{eq:J0Markov}) and (\ref{eq:CorrelationInvariant}). To do this, we need only demonstrate that the conservation law for the one-point function
\begin{equation}
	\partial_{t}\left(\mathcal{P}\u\mathcal{Q}_{h}\u\right)=-
	\partial_{h}\left(\mathcal{P}\left\{ -f(\u)\mathcal{Q}_{h}\u+\u\mathcal{Q}_{h}f(\u)\right\} \right)\label{eq:CorrelationConservationLaw2}
\end{equation}
holds. Recall that $
\partial_{h}\mathcal{Q}_{h}=\mathcal{A}\mathcal{Q}_{h}=\mathcal{Q}_{h}\mathcal{A}$. Then (\ref{eq:CorrelationConservationLaw2}) is equivalent to
\begin{equation}
	\mathcal{P}\left\{ f(\u)\mathcal{A}\mathcal{Q}_{h}\u-\u\mathcal{Q}_{h}\mathcal{A}f(\u)-\mathcal{B}\u\mathcal{Q}_{h}\u+\u[\mathcal{Q}_{h},\mathcal{B}]\u\right\} =0.\label{eq:CorrelationConservationLaw3}
\end{equation}
A short computation gives that
\begin{align*}
	f(\u)\mathcal{A}\mathcal{Q}_{h}\u-\u\mathcal{Q}_{h}\mathcal{A}f(\u)= & b(\u)f(\u)(
	\partial_{\u}\mathcal{Q}_{h}\u)-\u\mathcal{Q}_{h}(b(\u)f'(\u))\\
	& +f(\u)\int_{\mathbb{R}}n(\u,d\v)(\mathcal{Q}_{h}\v-\mathcal{Q}_{h}\u)\\
	& -\u\mathcal{Q}_{h}\left(\int_{\mathbb{R}}n(\u,d\v)(f(\v)-f(\u))\right).
\end{align*}
Similarly,
\begin{align*}
	\mathcal{B}\u\mathcal{Q}_{h}\u+\u[\mathcal{Q}_{h},\mathcal{B}]\u= & -b(\u)f'(\u)\mathcal{Q}_{h}\u-\u\mathcal{Q}_{h}(b(\u)f'(\u))\\
	& -\int_{\mathbb{R}}n(\u,d\v)(f(\v)-f(\u))\mathcal{Q}_{h}\v\\
	& -\u\mathcal{Q}_{h}\left(\int_{\mathbb{R}}n(\u,d\v)(f(\v)-f(\u))\right).
\end{align*}
Substituting these into (\ref{eq:CorrelationConservationLaw3}) and collecting terms yields (\ref{eq:CorrelationInvariant}).

Similar calculations hold for the $n$-point function and become clearer upon explicit use of the Manakov relation (\ref{eq:Manakov}). Using that constants are in the nullspace of $\mathcal{A}$ and $\mathcal{B}$, equations (\ref{eq:LaxEqn}) and (\ref{eq:MultiplicationOps}) imply that
\begin{align*}
	\partial_{t} & (\mathcal{P}\u\mathcal{Q}_{1}\u\cdots\mathcal{Q}_{n-1}\u)\\
	& =\mathcal{P}\mathcal{B}\u\mathcal{Q}_{1}\u\cdots\mathcal{Q}_{n-1}\u+\sum_{i=1}^{n-1}\mathcal{P}\u\mathcal{Q}_{1}\u\cdots\mathcal{Q}_{i-1}\u[\mathcal{Q}_{i},\mathcal{B}]\u\mathcal{Q}_{i+1}\u\cdots\mathcal{Q}_{n-1}\u\\
	& =\mathcal{P}[\mathcal{B},\mathcal{M}]\mathcal{Q}_{1}\u\cdots\mathcal{Q}_{n-1}\u+\sum_{i=1}^{n-1}\mathcal{P}\u\mathcal{Q}_{1}\u\cdots\mathcal{Q}_{i}[\mathcal{B},\mathcal{M}]\mathcal{Q}_{i+1}\u\cdots\mathcal{Q}_{n-1}\u.
\end{align*}
Substituting (\ref{eq:Manakov}), the previous expression becomes
\begin{align*}
	\mathcal{P}[\mathcal{N},\mathcal{A}]\mathcal{Q}_{1}\u\cdots\mathcal{Q}_{n-1}\u+\sum_{i=1}^{n-1}\mathcal{P}\u\mathcal{Q}_{1}\u\cdots\mathcal{Q}_{i-1}\u Q_{i}[\mathcal{N},\mathcal{A}]\mathcal{Q}_{i+1}\u\cdots\mathcal{Q}_{n-1}\u\\
	=\sum_{i=1}^{n-1}
	\partial_{h_{i}}\{\mathcal{P}\u\mathcal{Q}_{1}\u\cdots\mathcal{Q}_{i-1}(f(\u)\mathcal{Q}_{i}\u-\u\mathcal{Q}_{i}f(\u))\mathcal{Q}_{i+1}\u\cdots\mathcal{Q}_{n-1}\u\},
\end{align*}
which is equivalent to (\ref{eq:MarkovCorrelationConservationLaw}).

We remark that the above argument clarifies why quantities not of the form (\ref{eq:J0(n)Markov}), such as $\int\mathbb{E}[u^{2}(x)u(x+h)]dh$, are typically not conserved in time: 
\begin{align*}
	\partial_{t}\int_{0}^{\infty}\mathcal{P}\u^{2}\mathcal{Q}_{2}\u dh_{2} & =
	\partial_{t}\int_{0}^{\infty}\left.\mathcal{P}\u Q_{1}\u\mathcal{Q}_{2}\u\right|_{h_{1}=0}dh_{2}\\
	& =\int_{0}^{\infty}\mathcal{P}(f(\u)\mathcal{A}\u-\u\mathcal{A}f(\u))\mathcal{Q}_{2}\u dh_{2}\neq0
\end{align*}
except when $y$ and $f(y)$ are in the nullspace of $\mathcal{A}$.

\section{Exact solutions, energy, and spectra}

\subsection{White-noise initial data}

The invariant $J_{0}$ is expressed in terms of the 1- and 2-point functions for solutions which retain the Markov property in space. In certain special cases, its value can be explicitly obtained. For example, it is straightforward to compute $J_{0}$ when $u_{0}(x)$ is a stationary Ornstein-Uhlenbeck (mean-reverting) process:
\begin{equation}
	du_{0}(x)=-\beta u_{0}(x)dx+\gamma dB_{x},\qquad\beta,\gamma>0,\label{eq:OU}
\end{equation}
where $B_{x}$ is Brownian motion. Since $u_{0}$ has zero mean,
\begin{equation}
	\mathbb{E}[u_{0}(x)u_{0}(x+h)]=\text{Cov}(u_{0}(x),u_{0}(x+h))=\frac{\gamma^{2}}{2\beta}e^{-\beta h}.
\end{equation}
Therefore, $J_{0}=\gamma^{2}/(2\beta^{2})$.

By taking appropriate limits we obtain $J_{0}$ in the case of a white noise initial velocity field for any flux $f$. That is, suppose the initial potential is $U_{0}(x)=\sigma B_{x}$ with $\sigma>0$ the strength of the noise and $B_{x}$ a two-sided Brownian motion pinned at the origin. This corresponds to letting $\gamma=\sigma\beta$ and taking $\beta\to\infty$ in (\ref{eq:OU}), so that $J_{0}=\sigma^{2}/2$. The solution to Burgers equation with white noise initial data was obtained explicitly in terms of a Painlev\'e transcendent by Groeneboom \cite{Groeneboom}, and later rediscovered by Frachebourg and Martin \cite{Frachebourg}. The above computation thus determines the value of an integral of the form (\ref{eq:J0Markov}) with stationary and transition measures as in \cite{Frachebourg,Groeneboom}. For a more thorough examination of physically relevant quantities we refer the reader to \cite{V_whitenoise}.

\subsection{Energy dissipation and power spectrum}

$J_{0}$ is closely related to other quantities of physical interest, including the mean energy dissipation per unit interval and power spectrum. We now discuss some basic facts regarding these quantities for stationary Markov solutions to (\ref{eq:ConservationLaw}).

The mean dissipation in an interval $I$ is computed as follows. By Fubini's theorem and equation (\ref{eq:1-2PointEvolution}),
\begin{equation}
	\partial_{t}\mathbb{E}\left[\frac{1}{|I|}\int_{I}\frac{1}{2}u(x,t)^{2}dx\right]=\frac{1}{2}
	\partial_{t}\mathcal{P}\u^{2}=\frac{1}{2}\mathcal{P}\mathcal{B}\u^{2}.
\end{equation}
Let $h(\u)=\int_{\u_{0}}^{\u}\w f'(\w)d\w$ be the entropy flux associated to $\u^{2}/2$. Then 
\begin{align}
	\frac{1}{2}\mathcal{B}\u^{2} & =-b(\u)\u f'(\u)-\int_{\mathbb{R}}n(\u,d\v)[f]_{\u,\v}\left(\frac{\v^{2}}{2}-\frac{\u^{2}}{2}\right)\nonumber \\
	& =-\mathcal{A}h(\u)-\int_{\mathbb{R}}n(\u,d\v)\left\{ [f]_{\u,\v}\left(\frac{\v^{2}}{2}-\frac{\u^{2}}{2}\right)-(h(\v)-h(\u))\right\} .
\end{align}
Applying $\mathcal{P}$ and using that $\mathcal{P}\mathcal{A}h=0$, we obtain
\begin{align}
	\partial_{t} & \mathbb{E}\left[\frac{1}{|I|}\int_{I}\frac{1}{2}u(x,t)^{2}dx\right]\nonumber \\
	& =-\int_{\mathbb{R}}p(d\u)\int_{\mathbb{R}}n(\u,d\v)\left\{ \frac{1}{2}(\v+\u)(f(\v)-f(\u))+\int_{\v}^{\u}\w f'(\w)d\w\right\} .\label{eq:Dissipation}
\end{align}
This can also be arrived at by considering traveling wave solutions to (\ref{eq:ConservationLaw}) as in \cite{MP4}. In particular, the energy dissipated at any instant by a shock connecting the states $u_{-}>u_{+}$ is
\begin{align*}
	\frac{1}{2}(u_{+}+u_{-}) & (f(u_{+})-f(u_{-}))+\int_{u_{+}}^{u_{-}}wf'(w)dw\\
	& =\frac{1}{2}(u_{-}-u_{+})(f(u_{-})-f(u_{+}))-\int_{u_{+}}^{u_{-}}f(w)dw>0.
\end{align*}
Summing up over the expected number of shocks from $u_{-}$ to $u_{+}$ per unit interval implies (\ref{eq:Dissipation}).

Next, consider the power spectral density of the velocity field $u(x,t)$:
\begin{equation}
	E(k,t)=\frac{1}{2\pi}\int_{\mathbb{R}}e^{-ikh}\mathbb{E}[u(x,t)u(x+h,t)]dh,\qquad k\in\mathbb{R},t\geq0.
\end{equation}
Again, we drop $t$ from the notation. For stationary Markov solutions, the power spectrum is well-defined and can be given in terms of the $\lambda$-resolvent of the transition semigroup $(\mathcal{Q}_{h})_{h\geq0}$. With $\varphi$ a test function, let
\begin{equation}
	\mathcal{R}_{\lambda}\varphi(\u)=\int_{0}^{\infty}e^{-\lambda h}\mathcal{Q}_{h}\varphi(\u)dh,\qquad\text{Re }\lambda>0.
\end{equation}
Extending $\mathcal{R}_{\lambda}$ to $\lambda=ik$ with $0\neq k\in\mathbb{R}$, we have that
\begin{equation}
	E(k,t)=\frac{1}{\pi}\text{Re}\left(\mathcal{P}(t)\u\mathcal{R}_{ik}(t)\u\right).\label{eq:SpectrumResolvent}
\end{equation}
The Laplace transform of (\ref{eq:CorrelationConservationLaw2}) with respect to $h$ is therefore seen to be equivalent to an evolution equation for the power spectrum:
\begin{equation}
	\partial_{t}E=-\frac{k}{\pi}\text{Re}\left(\mathcal{P}\{f(\u)\mathcal{R}_{ik}\u-\u\mathcal{R}_{ik}f(\u)\}\right).
\end{equation}
To obtain $\mathcal{R}_{ik}$ in (\ref{eq:SpectrumResolvent}), one first needs to solve the Lax equation (\ref{eq:LaxEqn}). Finally, we note that $J_{0}=\lim_{k\to0}\pi E(k,t)$ is simply a constant multiple of the spectral density in the low-wavenumber limit.

Let us contrast the situation considered here with that of Burgers equation with one-sided L\'evy initial data as in \cite{MP4}. In the latter case, it was shown that there exist solutions with finite energy and infinite dissipation per unit interval due to an influx of energy through the boundaries. Here, the stationarity of the solution forbids such energy fluxes. It was also demonstrated in \cite{MP4} that the traditional notion of power spectrum is irrelevant when considering solutions with independent increments, as processes with the same energy have indistinguishable spectra. A useful notion of spectrum was instead given by the Fourier-Laplace transform of process paths without averaging. In contrast, here we have that the power spectrum is nontrivial and can be computed via the resolvent of the solution process.

\section{Acknowledgements}

I would like to thank Govind Menon for bringing this problem to my attention, and for the pleasure of a continuing collaboration. This work was partially funded by NSF grant DMS 06-36586.

\bibliographystyle{siam} 
\bibliography{s2}

\end{document}